\documentclass[a4paper,10pt]{article}

\usepackage[english]{babel}
\usepackage{graphicx}

\begin{document}
\begin{center}
\Huge Paper-and-pencil cosmological calculator \Large \\
\vskip0.5cm
\Large
S.V. Pilipenko\\
\textit{\normalsize Astrospace Center of the Lebedev Physical Institute, 117997 Moscow, Profsojuznaja 84/32}
\end{center}
\vskip 1cm
\large
A cosmological calculator is usually a program which computes relations between redshift, distance, physical and angular sizes, luminosity and apparent magnitude for some cosmological model characterized by a convenient set of parameters such as the Hubble constant at present, $H_0$, the dimensionless matter and dark energy densities $\Omega_m$ and $\Omega_\Lambda$. Several on-line calculators are available at NASA NED web site (http://ned.ipac.caltech.edu/help/cosmology\_calc.html). Although these calculators are ready to use, there are situations when there is no the Internet or a PC at hand. Online calculators also may be not enough vivid to use them in education since they usually calculate the answer only for a single input redshift. One solution to these two problems is a calculator based on the nomogram method.

The proposed paper-and-pencil calculator is designed for the $\Lambda$CDM cosmological model with recent cosmological parameters from the Planck mission: $H_0=67.4$ km/s/Mpc, $\Omega_\Lambda=0.685$ and $\Omega_m=0.315$ \cite{planck}. The calculator contains the following quantities:
\begin{itemize}
\large
\item z --- redshift;
\item H --- current value of the Hubble constant, km/s/Mpc;
\item r\_comov --- comoving distance, Mpc;
\item dm --- distance modulus;
\item age --- age of the Universe, Gyr;
\item time --- lookback time, Gyr;
\item size 1'' --- physical size of an object which is seen as an 1'' arc on the sky, kpc;
\item angle 1kpc --- angular size of a rod with physical size 1 kpc, arcsec.
\end{itemize}
In order to use the calculator, one needs to find a known value on a respective vertical scale. All the other values are situated on the same horizontal level. For the comfort of using a ruler, the redshift scale is repeated twice. The calculators are available for three redshift intervals: $z<20$, $z<1$, $z<0.1$. The space between major (labelled) tics on each vertical scale is always divided into ten equally-spaced intervals of the denoted value.

The code used to produce these calculators is public available and can be found at http://code.google.com/p/cosmonom/

\newpage
\topmargin=-4cm
\hspace{-3cm}
\includegraphics[]{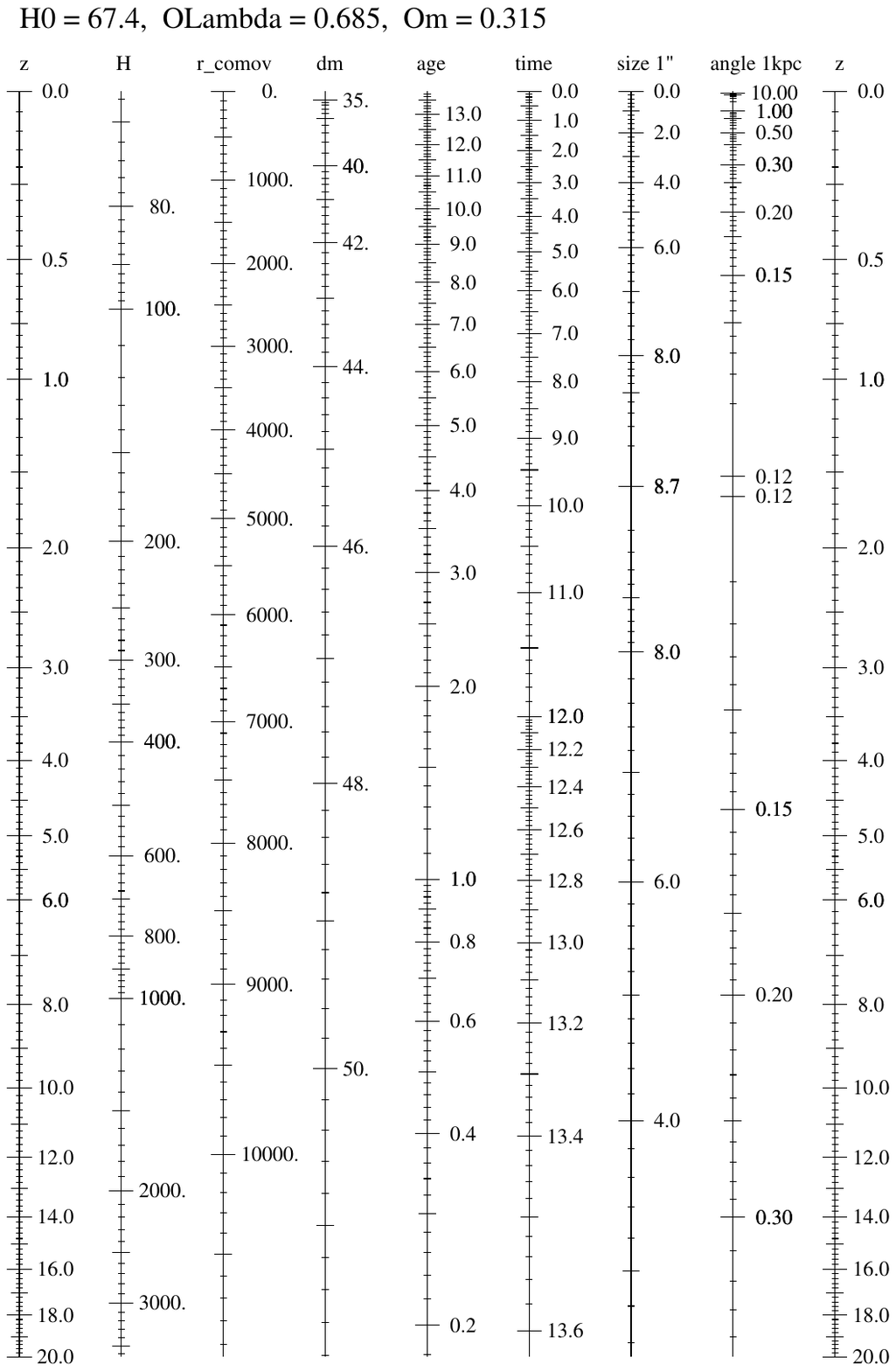}
\newpage
\hspace{-3cm}
\includegraphics[]{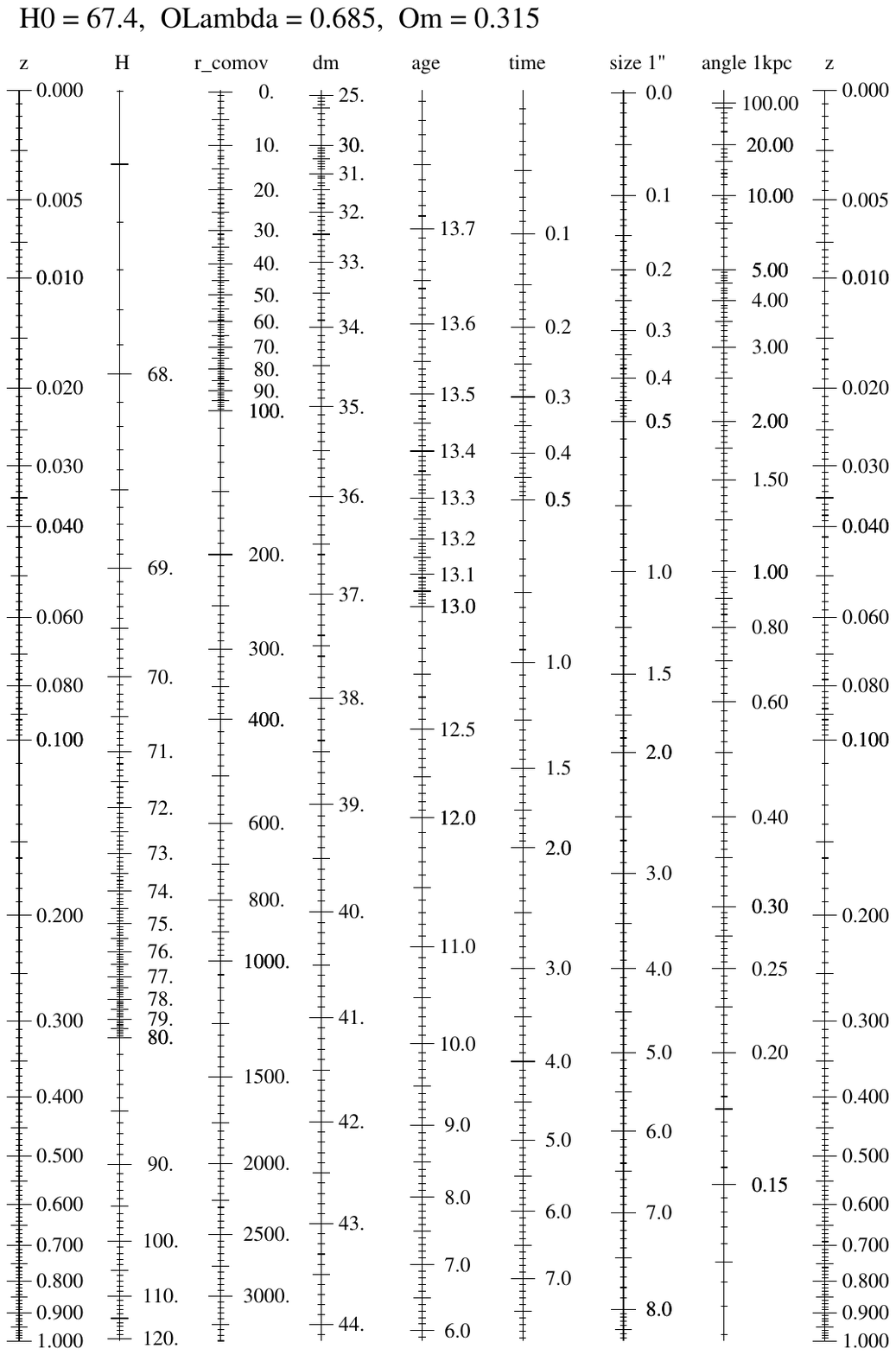}
\newpage
\hspace{-3cm}
\includegraphics[]{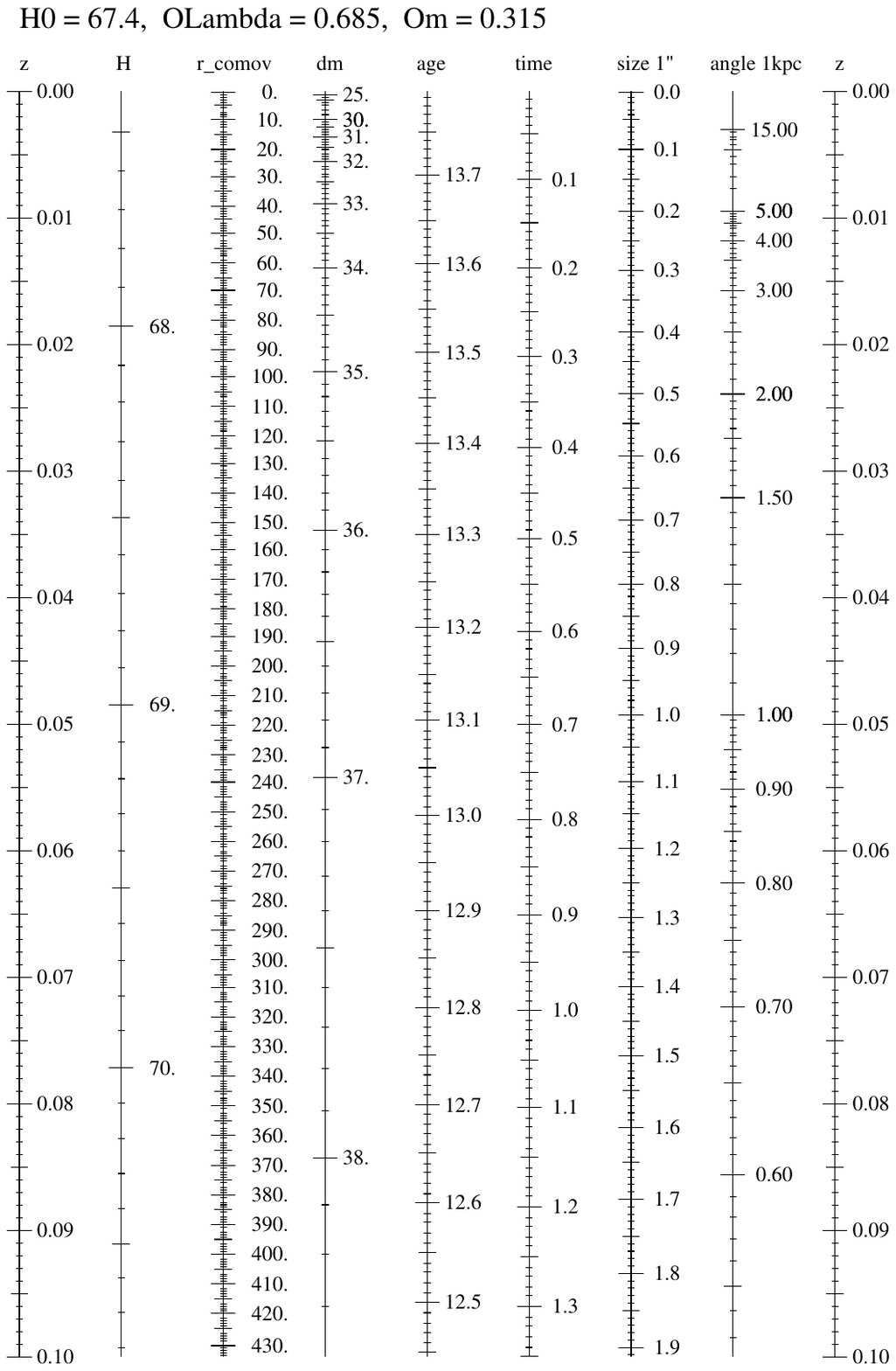}

\begin{thebibliography}{99}
\bibitem{planck}
Planck collaboration 2020 A\&A 641, A6, arXiv:astro-ph/1807.06209
\end{thebibliography}
\end{document}